\documentstyle[prl,aps,multicol]{revtex}

\begin{document}
\title{Frequency dependent relaxation rate in the superconducting YBa$_{2}$Cu$_{3}$O%
$_{6+\delta }$}
\author{A. Pimenov$^{1}$, A.Loidl$^{1}$, G. Jakob$^{2}$, and H. Adrian$^{2}$}
\address{$^{1}$Experimentalphysik V, EKM, Universit\"{a}t Augsburg, 86135 Augsburg,\\
Germany\\
$^{2}$Institut f\"{u}r Physik, Universit\"{a}t Mainz, 55099 Mainz, Germany}
\date{\today}
\maketitle

\begin{abstract}
The submillimeter-wave 3 cm$^{-1}$
\mbox{$<$}%
$\nu $\
\mbox{$<$}%
40 cm$^{-1}$ complex conductivity of the reduced YBa$_{2}$Cu$_{3}$O$%
_{6+\delta }$ film ($T_{\text{C}}$=56.5 K) was investigated for temperatures
4 K
\mbox{$<$}%
$T$
\mbox{$<$}%
300 K and compared to the properties of the same film in the optimally doped
state. The frequency dependence of the effective quasiparticle scattering
rate $1/\tau ^{*}(\nu )$ was extracted from the spectra. $1/\tau ^{*}$ is
shown to be frequency independent at low frequencies and high temperatures.
A gradual change to $1/\tau ^{*}\propto \nu ^{1.75\pm 0.3}$ law is observed
as temperature decreases. In order to explain the observed temperature
dependence of the low frequency spectral weight above $T_{\text{C}}$, the
quasiparticle effective mass is supposed to be temperature dependent for $T$%
\mbox{$>$}%
$T_{\text{C}}$.
\end{abstract}


\begin{multicols}{2}

It is now well established that the complex conductivity of high-$T_{\text{C}%
}$ superconductors has a highly unconventional character \cite{ir,tajima}.
Despite many experimental efforts an unresolved question remains: is the
frequency dependence of the conductivity due to a single mechanism or a sum
of completely different processes such as the Drude peak and the
mid-infrared absorption \cite{tanner}? Phenomenologically it has been proved
to be useful to present the conductivity data on the basis of the extended
Drude model with frequency-dependent effective mass $m^{*}$ and the
scattering rate $1/\tau ^{*}$ \cite{ir,pukhov}. The analysis of the infrared
conductivity has revealed a linear frequency dependence of the scattering
rate with a temperature-dependent offset value \cite{ir}. Complementary,
microwave techniques also allow the determination of the effective
scattering rate at much lower frequencies \cite{microwave}, and the
corresponding values can be considered to constitute the low-frequency limit
of the scattering rate. It is therefore reasonable to assume that $1/\tau
^{*}$ becomes nearly constant below some \ characteristic transition
frequency, as was recently observed \cite{bernhard} in the case of c-axis
conductivity in Y$_{1-x}$Ca$_{x}$Ba$_{2}$Cu$_{3}$O$_{6+\delta }$. Although
this assumption agrees well with most experimental data, it is extremely
difficult to observe this crossover frequency experimentally. This is mainly
due to the fact that both components of the complex conductivity $\sigma
^{*}=\sigma _{1}+i\sigma _{2}$ have to be measured with high accuracy for
frequencies below 50 cm$^{-1}$; a range which is rather difficult to explore
with conventional IR techniques. As the temperature decreases below $T_{%
\text{C}}$, the accurate determination of the scattering rate becomes even
more complicated because of the influence of the superconducting condensate
that has to be subtracted from the conductivity prior to the calculation of
the scattering rate. As a result, even less information exists concerning
the low-frequency behavior of the scattering rate below $T_{\text{C}}$.

In view of the problems outlined above, the method of \ the submillimeter
spectroscopy may be able to provide the necessary frequency-dependent
information. Using this method we have performed transmission experiments in
the frequency range 3 $\leq \nu \leq $ 40 cm$^{-1}$. The measurements were
carried out in a Mach-Zehnder interferometer arrangement \cite{volkov} which
allows both the measurements of transmission, and phase shift of a film on a
substrate. The properties of the blank substrate were determined in a
separate experiment. Utilizing the Fresnel optical formulas, the absolute
values of the complex conductivity can be determined directly from the
observed spectra without any approximations. Recently, using the
submillimeter spectrometer, we were able to obtain the complex conductivity
of the optimally doped YBa$_{2}$Cu$_{3}$O$_{6+\delta }$ film ($T_{\text{C}}$%
=89.5 K) and to directly observe the quasiparticle relaxation below $T_{%
\text{C}}$ \cite{recent}. In the present article we report on the properties
of the same film in an oxygen reduced state. Compared to the previous
experiments, a higher transmission value of the reduced sample allowed us to
estimate the frequency dependence of the effective scattering rate. In
addition, a broader temperature interval, in which both $\sigma _{1}(\nu )$
and $\sigma _{2}(\nu )$ are available, made it possible to determine the
spectral weight of the Drude peak and, thus, to observe the temperature
dependence of the effective mass of the quasiparticles.

The optimally doped 81 nm thick YBa$_{2}$Cu$_{3}$O$_{6+\delta }$ film on the
NdGaO$_{3}$ substrate \cite{recent} was oxygen depleted by heating up in a
controlled oxygen atmosphere, with subsequent quenching to room temperature.
X-ray diffraction showed the c-axis orientation of the film. The mosaic
spread of the c-axis oriented grains was 0.19$^{o}$. Four-point resistivity
measurement yielded an onset of the superconducting transition temperature, $%
T_{\text{C}}=$ 56.5$\pm 0.5$ K. The changes of the lattice constant and
critical temperature with oxygen depletion gave an estimate of the oxygen
content of the sample \cite{preparation}. For our sample a value of $\delta $%
=0.7$\pm $0.1 has been determined.

The frequency dependent scattering rate can be calculated from the complex
conductivity using the modified Drude expression \cite{ir,pukhov,allen}

\begin{equation}
\sigma _{D}^{*}=\varepsilon _{0}\omega _{p}^{2}(1/\tau ^{*}-i\omega )^{-1}
\end{equation}
where the plasma frequency $\omega _{p}$ and the renormalized scattering
rate $1/\tau ^{*}$ are assumed to be frequency dependent, $\omega =2\pi \nu $
is the circular frequency, and $\varepsilon _{0}$ is the permittivity of
free space. It should be noted that although the frequency dependences of $%
\omega _{p}${\it \ }and{\it \ }$1/\tau ^{*}$ are a parametrization only \cite
{hirschfeld,hensen}, the low frequency limit of these function can indeed be
connected to the quasi-particle self-energy \cite{hensen,shulga,thomas}.

Fig. 1 shows the frequency dependence of the complex conductivity of the YBa$%
_{2}$Cu$_{3}$O$_{6+\delta }$ film at different temperatures. The imaginary
part of the conductivity is presented in form of a product $\sigma _{2}\nu $%
, which allows the determination of the superconducting spectral weight via
\cite{jiang,carbotte}

\begin{equation}
\nu \sigma _{2}(\nu \rightarrow 0)=\frac{1}{2\pi }\frac{1}{\mu _{0}\lambda
(0)^{2}}=\frac{1}{2\pi }\varepsilon _{0}\omega _{p,s}^{2}
\end{equation}

In Eq. (2) $\lambda (0)$ represents the low-frequency limit of the
penetration depth, $\mu _{0}$ is the permeability of free space, and $\omega
_{p,s}^{2}$ is the spectral weight of the superconducting condensate. The
real part of the conductivity (lower frame of Fig.1) is frequency
independent at high temperatures and increases with the decreasing
temperature. At temperatures close to the superconducting phase transition a
frequency dependence of $\sigma _{1}(\nu )$ is observed which becomes
significant as the temperature is lowered further. At high temperatures the
imaginary part of the complex conductivity $\sigma _{2}\nu $ increases
approximately as $\nu ^{2}$. As the temperature decreases below $T_{\text{C}}
$, the low frequency offset of $\sigma _{2}\nu $ becomes nonzero, which
indicates the nonzero spectral weight of the superconducting condensate. The
overall frequency dependence of $\sigma _{2}\nu $ shows then a minimum at
zero frequency, with a characteristic width that becomes smaller with
decreasing temperature. In analogy to the complex conductivity data of the
unreduced YBa$_{2}$Cu$_{3}$O$_{6+\delta }$ sample \cite{recent} and
according to theoretical predictions \cite{carbotte}, the width of this
minimum qualitatively corresponds to the effective scattering rate.
Therefore, the temperature dependence of 1/$\tau ^{*}$ can be estimated
without any particular model assumptions. Nevertheless, in order to obtain
quantitative informations about the quasiparticle scattering and spectral
weight, the conductivity was analyzed using the simple Drude model. The
effects arising from the superconducting component were taken into account
by adding an appropriate term $\sigma _{s}$ to the imaginary part of
conductivity \cite{bonn93}. The final expression for nonzero frequencies can
then be written in the form

\begin{equation}
\sigma ^{*}(\omega )=\sigma _{D}^{*}+\sigma _{s}==\varepsilon _{0}\omega
_{p,n}^{2}(1/\tau ^{*}-i\omega )^{-1}+i\varepsilon _{0}\omega
_{p,s}^{2}/\omega  \label{drudesup}
\end{equation}
where $\omega _{p,n}^{2}$, $\omega _{p,s}^{2}$ and $1/\tau ^{*}$ \ are
frequency independent, $\omega _{p,s}^{2}$ represents the spectral weight of
the superconducting condensate (Eq. 2), $\omega _{p,n}^{2}$ is the spectral
weight of the nonsuperconducting component, and $1/\tau ^{*}$ is the
characteristic scattering rate. The results of the simultaneous fitting of
the real and the imaginary parts of conductivity are represented as solid
lines in Fig. 1 and provide reasonable description of the low-frequency
experimental data. Prominent deviations are observed below $T_{\text{C}}$
and at high frequencies. The experimental data have a significantly weaker
frequency dependence compared to the Drude model. Therefore, a frequency
dependent scattering rate has to be used in order to fit the experimental
data correctly. The deviations become less apparent in $\sigma _{2}\nu $ at
low temperatures due to the dominance of the superconducting condensate ($%
\sigma _{s}$ in Eq. 3).

In order to obtain the frequency dependence of the scattering rate directly
from the complex conductivity, we recalculated $1/\tau ^{*}(\nu )$ from the
experimental data using Eq. (1): $1/\tau ^{*}=\omega \sigma _{1}(\omega
)/\sigma _{2}(\omega )$. The procedure is quite straightforward for {\em T}$%
> ${\em T}$_{\text{C}}$ because $\omega _{p,s}$=0. For {\em T}$\leq ${\em T}$%
_{\text{C}}$ the term $\sigma _{s}$, describing the effect of
superconducting condensate, has to be subtracted from the imaginary part of
the conductivity. The obtained frequency dependence of the effective
scattering rate is presented in the Fig. 2. The scattering rate is almost
frequency independent at high temperatures in the submillimeter frequency
range, as is well documented by the 80 K data set. A significant frequency
dependence evolves below T$_{C}$ with a crossover to a constant value. The
crossover frequency shifts to lower frequencies as the temperature
decreases. At $T$=6 K the scattering rate reveals a single power-law
behavior and can be approximated by $1/\tau ^{*}\propto \nu ^{1.75\pm 0.3}
$. This power law is close to the $\nu ^{2}$ behavior, which was observed in
the normal state and at infrared frequencies in reduced YBa$_{2}$Cu$_{3}$O$%
_{6+\delta }$ sample \cite{basov96}; however it is not clear, whether the
same processes are determining the low frequency electrodynamics above and
below $T_{\text{C}}$.

Figure 3 shows the temperature dependence of the scattering rate of reduced (%
$T_{\text{C}}$ = 56.5 K) and optimally doped ($T_{\text{C}}$ = 89.5 K) films
as determined using the simple Drude analysis (Eq. (3)) of the complex
conductivity. These data may be considered as the limiting low-frequency
value of the scattering rate. The dotted lines in the insert indicate the
possible linear temperature dependence of $1/\tau ^{*}(T)$. Recently, Ioffe
and Millis \cite{millis} proposed a model which imply a quadratic rather
than linear temperature dependence of the scattering rate above $T_{\text{C}%
}.$ Unfortunately, the present data do not allow a direct proof of this
dependence, as the scattering rate above 150 K for both samples has been
obtained assuming a temperature independent Drude spectral weight $\omega
_{p,n}^{2}=const(T>150K)$. As will be seen below, this assumption may have
substantial influence on the form of the $1/\tau ^{*}(T)\,$curve. Below 150
K the experimental results for the underdoped sample start to deviate from
the dotted line. This effect is not observed for the optimally doped sample
and can be attributed to the opening of the spin gap in the reduced sample
\cite{uchida} below a characteristic temperature $T^{*}>T_{\text{C}}$.

Figure 4 shows the temperature dependence of the Drude spectral weight for
the optimally doped (left frame) and reduced (right frame) samples. The
absolute value of the total spectral weight for the oxygen reduced sample is
lower compared to the optimally doped sample for all temperatures. This
agrees well with the doping dependence of the total spectral weight as
obtained by infrared measurements in the normal state \cite{rotter} and by
the magnetic penetration depth measurements in the superconducting state
\cite{lambda,uemura}. Fig. 4 shows a strong temperature dependence of the
spectral weight of a reduced sample above $T_{\text{C}}$. A possible
explanation of this temperature dependence can be given in terms of a
temperature dependence of the effective mass of the quasiparticles, because
the Drude spectral weight may be written as $\varepsilon _{0}\omega
_{p,n}^{2}=ne^{2}/m^{*}$. According to the kinetic inductance data of Fiory
et al. \cite{fiory}, the quasiparticle concentration remains temperature
independent above $T_{\text{C}}$. Therefore, our data suggest an increase of
the effective mass of quasiparticles at low frequencies, approximately a
factor of three, as the temperature is lowered from 150 to 60 K. These
results may be compared with the two-component analysis of the infrared
conductivity \cite{ir} carried out for frequencies above 50 cm$^{-1}$. This
analysis revealed a nearly temperature independent weight of the Drude
component for optimally doped YBa$_{2}$Cu$_{3}$O$_{6+\delta }$ \cite
{ir,infrared}. On the other hand, the analysis of the infrared conductivity
on the basis of the modified Drude model (i.e. one-component analysis)
reveal a remarkable temperature dependence of the effective mass in the low
frequency limit \cite{ir,pukhov,thomas} both for underdoped and for
optimally doped cuprates. Since both types of analysis are expected to
coincide in the low frequency limit \cite{pukhov}, the two results
apparently contradict each other. Interestingly, our submillimeter data of
the spectral weight are obtained by the simple Drude analysis of the first
type, but provide the temperature dependent effective mass similar to the
low frequency limit of the modified Drude analysis.

As the temperature is lowered through $T_{\text{C}}$, the normal spectral
weight ($\omega _{p,n}^{2}$, Fig. 4) for both films decreases and then
saturates at a finite value even at $T$%
\mbox{$<$}%
0.1 $T_{\text{C}}$. This decrease is followed by a gradual increase of the
spectral weight of the superconducting component. As a result the full
spectral weight for both samples reveals almost no changes for $T$%
\mbox{$<$}%
$T_{\text{C}}$ compared to $\omega _{p,s}^{2}+\omega _{p,n}^{2}$ at $T$=$T_{%
\text{C}}$. This indicates that the apparent temperature dependence of the
effective mass is ''frozen'' below $T_{\text{C}}$. Therefore, the two-fluid
model assumption, $n_{n}+n_{s}=const$, which supposes a temperature
independent effective mass, holds for $T\leq T_{\text{C}}$. For higher
temperatures, the conservation of the low frequency spectral weight is
violated due to the temperature dependent $m^{*}$.

In summary, we have investigated the submillimeter-wave complex conductivity
of a reduced YBa$_{2}$Cu$_{3}$O$_{6+\delta }$ film ($T_{\text{C}}$=56.5 K)
and compared it to the properties of the same film in the optimally doped
state. Higher transparency and lower transition temperature of the reduced
film allowed the observation of qualitatively new effects. The frequency
dependence of the effective quasiparticle scattering rate has been extracted
from the conductivity spectra. It was possible to show experimentally that
the scattering rate is frequency independent at low frequencies and high
temperatures. For decreasing temperature a transition between $1/\tau
^{*}=const$ and $1/\tau ^{*}\propto \nu ^{1.75\pm 0.3}$ is observed. In
addition, the low frequency spectral weight of the Drude component was
estimated as a function of temperature and is shown to be temperature
dependent above $T_{\text{C}}$. In order to explain the observed behavior,
one has to assume an increase of the effective quasiparticle mass by a
factor of three as the temperature is lowered from 150 K to 60 K. On the
contrary, the total low frequency spectral weight $\omega _{p,n}^{2}+\omega
_{p,s}^{2}$ is temperature independent for $T\leq T_{\text{C}}$.

We thank K. Scharnberg, J. K\"{o}tzler, A. Kampf and K. Held for valuable
discussions. This work was supported by the BMBF via the contract number
13N6917/0 - Elektronische Korrelationen und Magnetismus.

\bigskip
Figure captions.

\medskip

Fig. 1. Frequency dependence of the complex conductivity of the reduced YBa$%
_{2}$Cu$_{3}$O$_{6+\delta }$ film at different temperatures. Upper panel:
the product $\sigma _{2}\nu $; lower panel: $\sigma _{1}$. Solid lines are
fits according to Eq. (3). Dotted lines are drawn to guide the eye. Arrows
indicate the approximate positions of the quasiparticle relaxation.

\smallskip

Fig. 2. Frequency dependence of the quasiparticle scattering rate of the
reduced YBa$_{2}$Cu$_{3}$O$_{6+\delta }$ film at different temperatures.
Solid lines are guides to the eye. Arrows at low frequencies indicate the
T=10 K values from Ref. \cite{bonn94}.

\smallskip

Fig. 3. Temperature dependence of the effective scattering rate of the
reduced ($T_{\text{C}}$=56.5 K) and the optimally doped ($T_{\text{C}}$=89.5
K, \cite{recent}) YBa$_{2}$Cu$_{3}$O$_{6+\delta }$ film as extracted from
the fits (Eq. 3)\ to the complex conductivity data. The insert shows the
data on the linear scale. Solid lines are guide to the eye. Dashed lines
represent an extrapolation of the linear temperature dependence of the
scattering rate observed at high temperatures.

\smallskip

Fig. 4. Temperature dependence of the effective low frequency spectral
weight of the reduced ($T_{\text{C}}$=56.5 K, right frame) and the optimally
doped ($T_{\text{C}}$=89.5 K, left frame, \cite{recent}) YBa$_{2}$Cu$_{3}$O$%
_{6+\delta }$ film. The data are extracted from the low frequency offset of $%
\sigma _{2}\nu $ (superconducting component, $\omega _{p,s}^{2}$) and from
the Drude fits (Eq. 3) to the complex conductivity data (normal component, $%
\omega _{p,n}^{2}$).

\end{multicols}

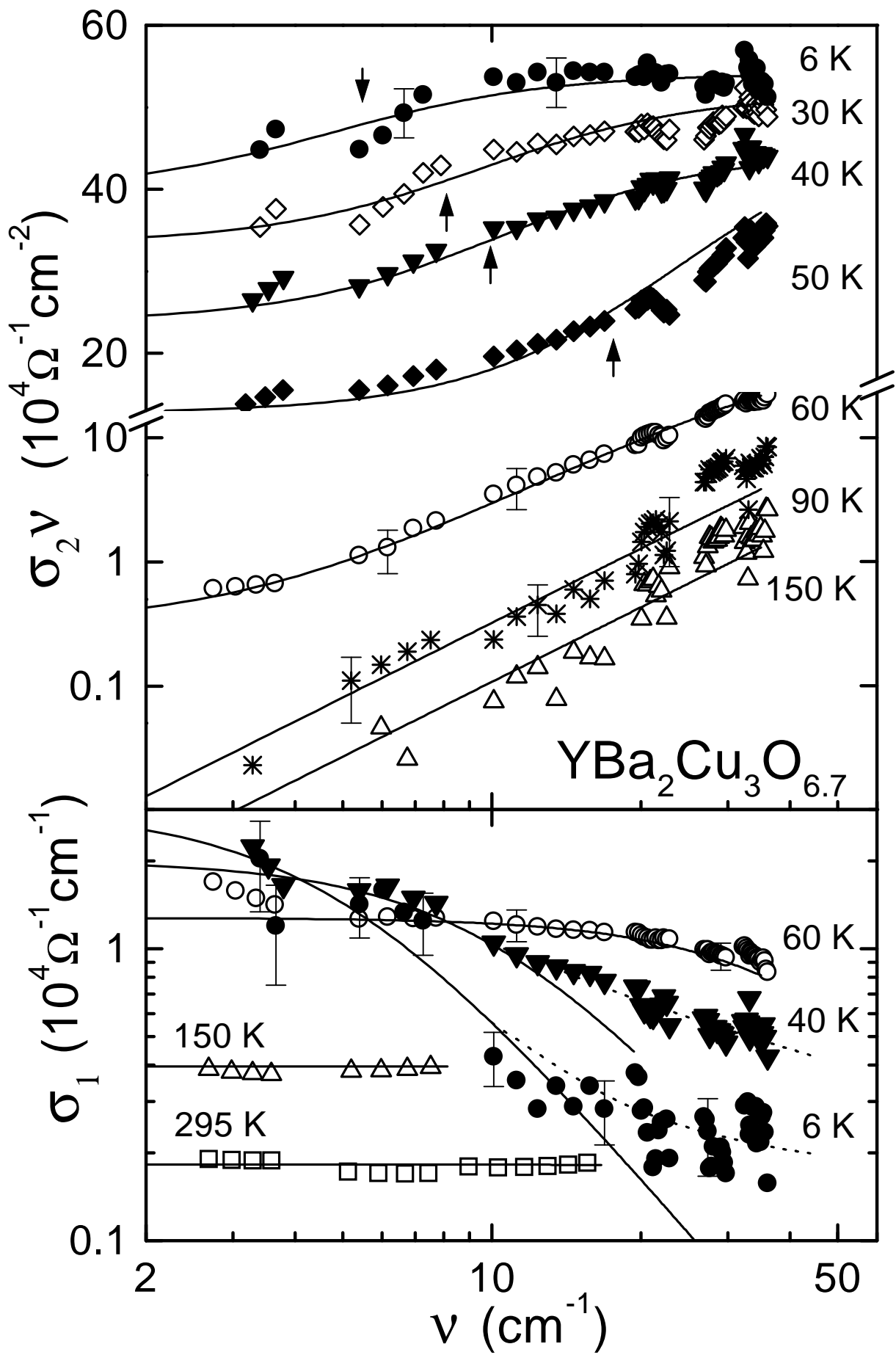
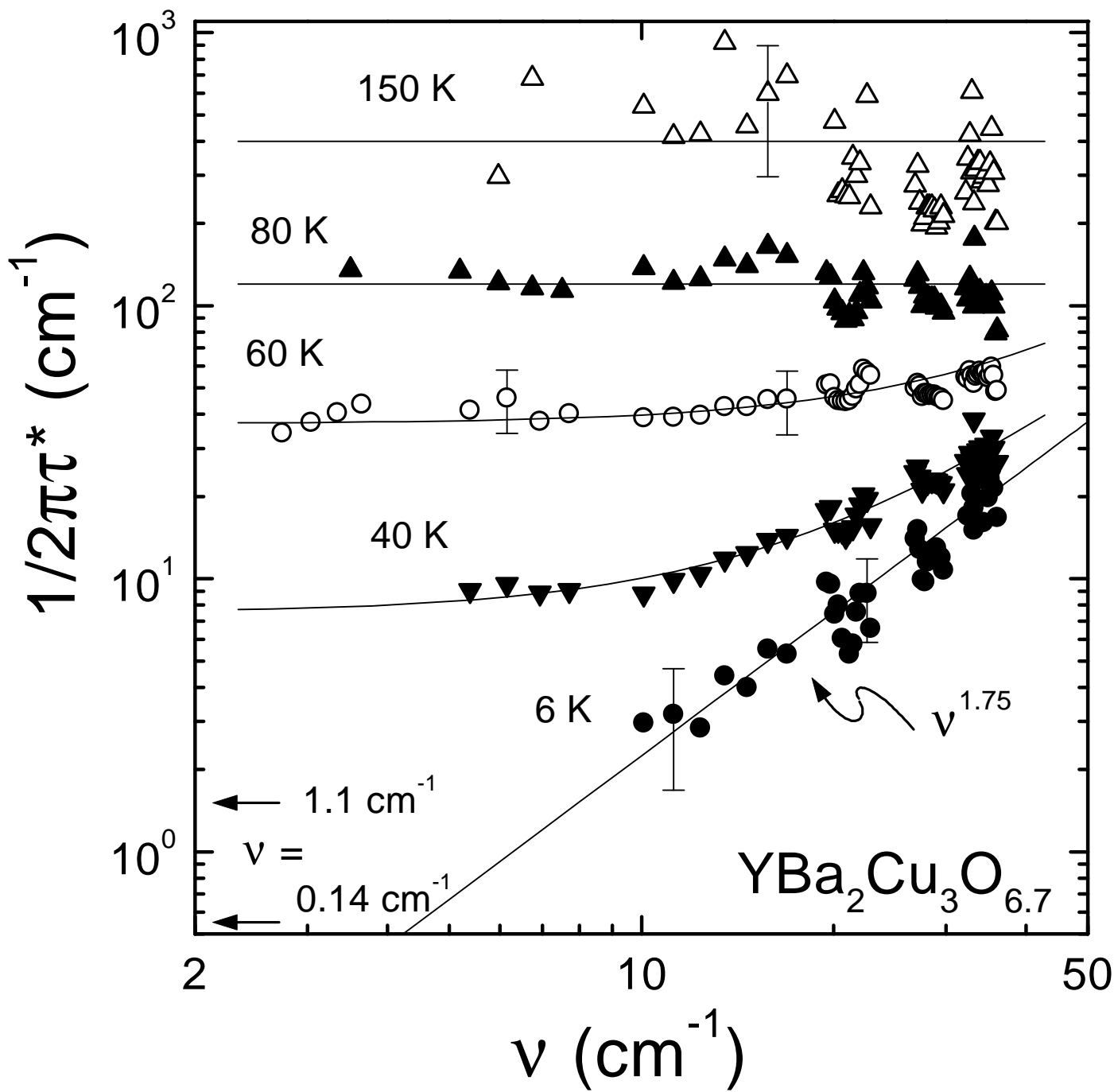
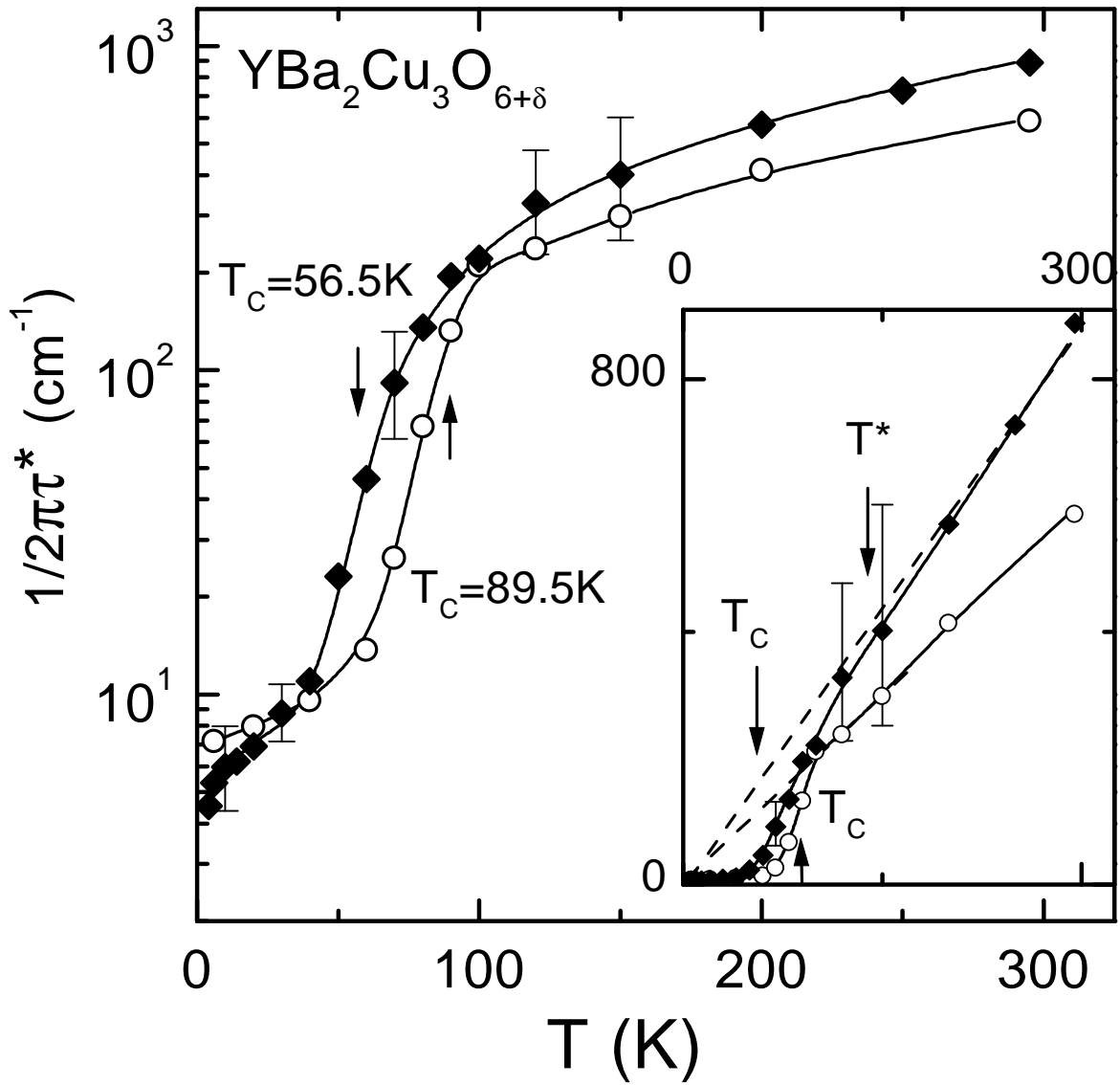
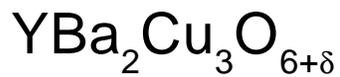
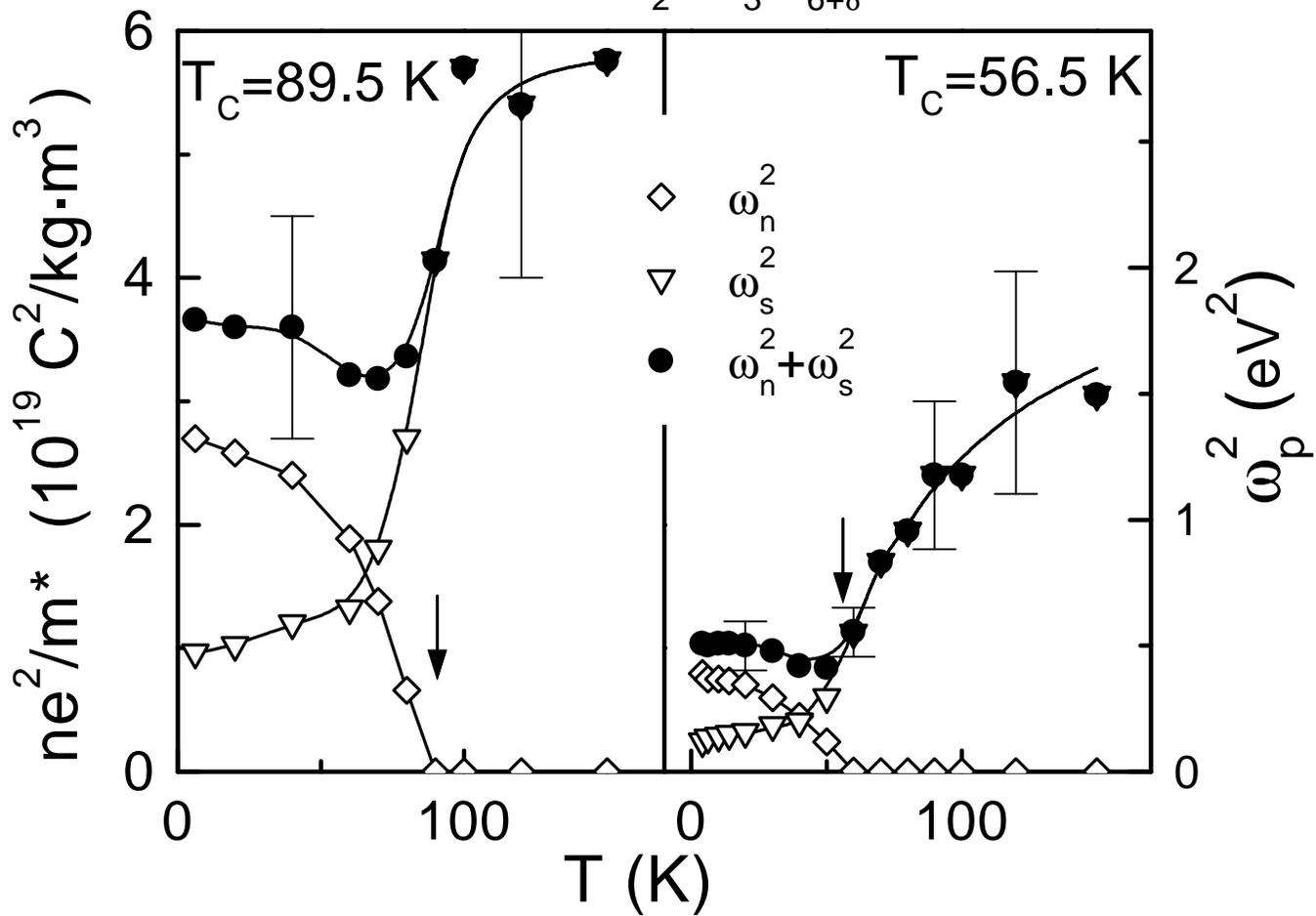
\end{document}